\documentclass[10pt,conference]{IEEEtran}

\usepackage[tbtags]{amsmath}
\usepackage{amsfonts}
\usepackage{amssymb}
\usepackage{subfigure}
\usepackage{cite}
\usepackage{calc}
\usepackage{color}
\usepackage{epsfig}

\newtheorem{defn}{Definition}
\newtheorem{thm}{Theorem}[section]
\newtheorem{cor}[thm]{Corollary}
\newtheorem{prop}{Proposition}
\newtheorem{lem}[thm]{Lemma}
\newtheorem{conj}[thm]{Conjecture}
\newtheorem{constr}[thm]{Construction}
\newtheorem{note}{Remark}
\newcommand{\bit}{\begin{itemize}}
\newcommand{\eit}{\end{itemize}}
\newcommand{\bcor}{\begin{cor}}
\newcommand{\ecor}{\end{cor}}
\newcommand{\beq}{\begin{equation}}
\newcommand{\eeq}{\end{equation}}
\newcommand{\beqn}{\begin{equation*}}
\newcommand{\eeqn}{\end{equation*}}
\newcommand{\beqa}{\begin{eqnarray}}
\newcommand{\eeqa}{\end{eqnarray}}
\newcommand{\beqan}{\begin{eqnarray*}}
\newcommand{\eeqan}{\end{eqnarray*}}
\newcommand{\ben}{\begin{enumerate}}
\newcommand{\een}{\end{enumerate}}
\newcommand{\bdefn}{\begin{defn}}
\newcommand{\edefn}{\end{defn}}
\newcommand{\bnote}{\begin{note}}
\newcommand{\enote}{\end{note}}
\newcommand{\bprop}{\begin{prop}}
\newcommand{\eprop}{\end{prop}}
\newcommand{\blem}{\begin{lem}}
\newcommand{\elem}{\end{lem}}
\newcommand{\bthm}{\begin{thm}}
\newcommand{\ethm}{\end{thm}}
\newcommand{\bconj}{\begin{conj}}
\newcommand{\econj}{\end{conj}}
\newcommand{\bconstr}{\begin{constr}}
\newcommand{\econstr}{\end{constr}}
\newcommand{\bpf}{\begin{proof}}
\newcommand{\epf}{\end{proof}}

\begin{document}

\title{Diversity and Degrees of Freedom of Cooperative Wireless Networks}
\author{K. Sreeram, S. Birenjith,  P. Vijay Kumar}

\author{\authorblockN{K. Sreeram}
\authorblockA{Department of ECE\\
Indian Institute of Science\\
Bangalore, India \\
Email: sreeramkannan@ece.iisc.ernet.in} \and
\authorblockN{S. Birenjith}
\authorblockA{Department of ECE\\
Indian Institute of Science\\
Bangalore, India \\
Email: biren@ece.iisc.ernet.in} \and
\authorblockN{P. Vijay Kumar}
\authorblockA{Department of ECE\\
Indian Institute of Science\\
(on leave from USC)\\
Email: vijayk@usc.edu}}

\date{}
\maketitle

\begin{abstract}

Wireless fading networks with multiple antennas are typically
studied information-theoretically from two different perspectives
- the outage characterization and the ergodic capacity
characterization. A key parameter in the outage characterization
of a network is the diversity, whereas a first-order indicator for
the ergodic capacity is the degrees of freedom (DOF), which is the
pre-log coefficient in the capacity expression. In this paper, we
present max-flow min-cut type theorems for computing both the
diversity and the degrees of freedom of arbitrary single-source
single-sink multi-antenna networks. We also show that an
amplify-and-forward protocol is sufficient to achieve this. The
degrees of freedom characterization is obtained using a conversion
to a deterministic wireless network for which the capacity was
recently found. We show that the diversity result easily extends
to multi-source multi-sink networks and evaluate the DOF for
multi-casting in single-source multi-sink networks.

\end{abstract}

\section{Introduction}

There has been a recent interest in determining the degrees of
freedom (DOF) of wireless multi-antenna networks \cite{HostNosr}.
For definitions of diversity and degrees of freedom of
point-to-point channels, see \cite{GodHero} for example. The DOF
for the $N$ user interference channel was recently derived in
\cite{CadJaf} and the DOF of single-source single-sink layered
networks was obtained in \cite{BorZheGal}.

We characterize the diversity for arbitrary networks and compute degrees of freedom (DOF) for single-source
single-sink and multi-cast networks with multiple antennas. We compute the degrees of freedom using a connection
to deterministic wireless networks. The capacity of single-source single-sink and multi-cast deterministic
wireless networks were characterized in \cite{ADT1}. Intuition drawn from the deterministic wireless networks
were used to identify capacity to within a constant for some example networks in \cite{ADT2}. A similar approach
was used in \cite{CadJafSha} for obtaining DOF for real gaussian interference networks.

While the results for wire-line finite-field single-source
single-sink network have been known for some time \cite{FordFulk},
multi-cast capacity was found in the more recent seminal work
\cite{NetInfFlow}. An algebraic approach for finding the
multi-cast capacity was given in \cite{KotMed}. These results were
extended to finite field wireless networks in \cite{ADT1}. In
\cite{NazGast}, computation codes were used to study the capacity
of finite field networks with interference alone. While it is easy
to extend wireline finite field network results to gaussian
wireline networks, the extension of wireless finite field network
results to the gaussian case is not obvious. In this paper, we
apply these finite field network results to compute the DOF and
diversity for gaussian wireless networks. Table~\ref{tab:net_cod}
summarizes these developments.

The diversity of a family of multi-hop networks was evaluated in \cite{YanBelNew}. In \cite{YanBelMimoAf}, the
diversity for two-hop MIMO relay channel with a certain condition on the number of antennas. However the maximum
diversity of an arbitrary network remains an open question, which we settle in this paper.

\begin{table}
\caption{Network Coding for Finite Field and Gaussian Networks \label{tab:net_cod}}
\begin{center}
\begin{tabular}{|c|c|c|c|c|}
\hline
&\multicolumn{2}{|c|}{}&\multicolumn{2}{|c|}{}\\
%Hey & \multicolumn{2}{|c|}{Team sheet} \\
&\multicolumn{2}{|c|}{Wireline Networks}&\multicolumn{2}{|c|}{Wireless Networks}\\
\hline
&&&&\\
&Capacity of&DOF of&Capacity of&DOF of\\
&Finite Field &Gaussian&Finite Field&Gaussian\\
&Networks &Networks&Networks&Networks\\
\hline
&&&&\\
Single & Min-cut & Min-cut & Min-cut & Min-cut\\
Source & \cite{FordFulk} & (easy & rank  \cite{ADT1}&  rank\\
&&to see) && (this paper)\\
\hline
&&&&\\
Multicast & Minimum & Minimum & Minimum & Minimum \\
&  min-cut \cite{NetInfFlow}& min-cut &min-cut  & min-cut rank\\
&&(easy to see) &rank \cite{ADT1} & (this paper)\\
\hline
\end{tabular}
\end{center}
\end{table}

\subsection{Representation of a Network} We represent a single-antenna wireless
network by a edge labelled directed graph $\mathcal{G} = (\mathcal{V},\mathcal{E}
)$, where $\mathcal{V}$ is the set of vertices and $\mathcal{E}$ is the set of
edges. Each node is represented by a vertex, each edge represents a transmission
link. Let $N := |\mathcal{E}|$ be the number of links in the network. The label on
every edge $\mathcal{L}(e), e \in \mathcal{E}$ represents the fading coefficient on
that transmission link. By convention, we put an edge only when the link has
non-zero fading coefficient.

In the case of multiple antenna networks, we first pass on to an
equivalent representation, where every terminal is represented by
a super-node and every antenna attached to the terminal is
represented by a small node associated with the super-node. Edges
representing single-antenna connections are drawn only between
small nodes and hence we can still label edges by scalar fading
coefficients. For the gaussian network, we assume that
coefficients are elements of the complex field $\mathbb{C}$. Since
we are dealing with wireless networks, we assume that the
broadcast and interference constraints hold. We assume throughout
the paper that CSIR is present. We also assume for the degrees of
freedom part, that CSIT is present.

\bdefn A cut $\omega$ between source $S_i$ and destination $D_j$
on a multiple-antenna gaussian network is defined as a partition
of the super-nodes into ${U}$ and ${U}^c$ such that the source
$S_i$ is present in ${U}$ and $D_j$ is present in ${U^c}$. Let the
set of all cuts between source $S_i$ and destination $D_j$ be
denoted by $\Omega_{ij}$. Given a cut $\omega$, the matrix of the
cut, $H_{\omega}$ is defined as the transfer matrix associated
with edges crossing the cut from the source side to destination
side. In the single source single sink case, we will drop the
unneeded ${ij}$ suffix.\edefn

\bnote Any wire-line network can be converted into a wireless
network, by adding a sufficient number of small-nodes at each
super-node thereby separating the links so that in effect, the
broadcast and interference constraints are nullified. We call this
the natural embedding of a wire-line network into a wireless
network. \enote

\subsection{Degrees of Freedom}

\bdefn Consider a single-sink wireless network with each node having a symmetric
transmit power constraint, $\rho$. Let the capacity between the source $i$ and the
sink $j$ be $C_{ij}(\rho)$. The \emph{degrees of freedom} of the flow between source
$i$ and sink $j$ is defined as
 \beqa D_{ij} & = & \lim_{\rho \rightarrow \infty} \frac{C_{ij}(\rho)}{\log \rho} \eeqa \edefn

\vspace{0.1in} \bnote The capacity of the flow $F_{ij}$ between
can source $i$ and sink $j$ can then be given by: \beqa C_{ij} & =
& D_{ij} \log(\rho) + o(\log \rho) \eeqa Whenever we consider a
single-source single-sink network, we will without loss of
ambiguity, drop the suffix from $C_{ij}, D_{ij}$ and simply use
$C,D$ instead. \enote

\vspace{0.1in} \bdefn A \emph{multi-cast} network is defined as a network with
single-source and multiple-sinks, with the constraint that all the flows need the
same information from the source. \edefn

\blem \label{lem:DOF_rank}Consider a channel of the form $\bold{Y
= HX + W}$, where $H$ is a $N \times N$ channel matrix $X,Y,W$ are
$N$ length column vectors representing the transmitted signal,
received signal and the noise vector distributed as
$\mathcal{CN}(0,\Sigma)$, where $\Sigma$ is a full rank
correlation matrix. The degrees of freedom of this channel is
given by $D = \text{rank} (H)$. \elem

%\bpf Consider the singular value decomposition (svd) of $H$ into the form $U \Lambda
%V^\dagger$, where $U,V$ are unitary matrices and $\Lambda$ is a diagonal matrix
%comprised of the singular values. Clearly, the capacity of the channel $H$ depends
%only on $\Lambda$, since capacity is given by
%
%\beqan C & = & \log \det (I + \Sigma_xHH^\dagger) \\
% & = & \log \det (I + \Sigma_x\Lambda {\Lambda}^\dagger) \eeqan
%where $\Sigma_x$ is the correlation matrix of $X$.
%
%Let $\lambda_i, i=1,2,...,n$ be the singular values of the matrix. Let
%$\lambda_{\text{min}}$,$\lambda_{\text{max}}$ be the singular values with the
%minimum and maximum absolute value. Now
%
%\beqan |\lambda_{\text{min}}|^2 I \ \preceq & \Lambda\Lambda^\dagger & \preceq
%|\lambda_{\text{max}}|^2 I \eeqan
%
%Now since the function $\log \det$ is non-decreasing on the cone of positive
%semi-definite matrices, we have that
%
%\beqan \log \det ( I + \rho \Sigma_x |\lambda_{\text{min}}|^2 I) & \leq & \log \det
%( I + \rho \Sigma_x \Lambda\Lambda^\dagger ) \\
%&  \leq  & \log \det ( I + \rho \Sigma_x |\lambda_{\text{max}}|^2 I) \eeqan
%
%Therefore we are able to bound the DOF by the DOF of the two matrices
%$\lambda_{\text{min}}I$ and $\lambda_{\text{max}}I$. Since both matrices have DOF
%equal to $rank (H)$, we get that DOF of the channel stated in the Lemma is rank (H).
% \epf

\bnote We define the DOF of a matrix $H$ as the DOF of the channel
$Y=HX+W$ where $W$ is $\mathcal{CN}(0,I)$. \enote

\subsection{Diversity}
The Lemma below computes the diversity of a channel matrix having
a specific structure.

\blem \label{lem:Diversity}Consider a channel of the form $\bold{Y
= HX + W}$, where $H$ is a $N \times N$ random channel matrix,
$X,Y,W$ are $N$ length column vectors representing the transmitted
signal, received signal and the noise vector. Let the noise vector
$W$ can be representable as $W = Z + Z_0 + \sum_{i=1}^{L} G_iZ_i$,
where $Z_i$ are $\mathcal{CN}(0,I)$ independent vectors and every
entry in the matrices $G_i$ are polynomial functions of gaussian
random variables. If the $N^2$ entries of matrix $H$ contain
exactly $M$ independent Rayleigh fading coefficients, then $M$ is
the diversity of that matrix. \elem

\subsection{Linear Deterministic Wireless Network}

 In defining
deterministic\footnote{By deterministic network, we will always
mean linear deterministic network.} wireless networks, we follow
\cite{ADT1}.  These networks are similar to multiple-antenna
gaussian networks with the only difference being that these
networks are noise-free and that the complex fading coefficients
are replaced by finite fields elements.
% A deterministic wireless network is
%defined via an edge-labelled graph as in Gaussian wireless
%network. However, the entire network is assumed to be operating
%over a fixed finite field $F_p$ and noise is absent in the
%network.
In place of complex vectors, each node transmits an $q$-tuple over
 the finite field.
Cuts are defined as in the gaussian network case. In place of
$H_{\omega}$, we use $G_{\omega}$ to define the transfer matrix
between nodes on either side of the cut $\omega$. We state the
following Theorem from \cite{ADT1}:

\bthm \cite{ADT1} \label{thm:ADT_ss} Given a linear deterministic
single-source single-sink wireless network over any finite field
$\mathbb{F}$, $\forall \ \epsilon
> 0$,  the $\epsilon$-error capacity $C$ of such a relay network
is given by,
\begin{eqnarray}
\label{eq:ThmLinDetNet} C = \min_{\omega \ \in \ \Omega} \mathrm{rank}(G_{\omega}).
\end{eqnarray}
where the capacity is specified in terms of the number of finite
field symbols per unit time. A strategy utilizing only linear
transformations over $\mathbb{F}$ at the relays is sufficient to
achieve this capacity. \ethm

\bnote The strategy specified in \cite{ADT1} utilizes matrix
transformations at each relay of the input vector received over a
period of $T$ time slots. Thus the achievability shows the
existence of relay matrices $A_i$ at each relay node $i \in
\mathcal{V}$, each of size $qT \times qT$, that specifies the
transformation between the received vector of size $qT$ to the
vector of size $qT$ that needs to be transmitted. It can be seen
using the natural embedding of a wire-line network into a wireless
network, that this theorem is indeed a generalization of the
max-flow min-cut theorem.  \enote The multicast-version of
Theorem\ref{thm:ADT_ss} appears below.

\bthm \cite{ADT1} \label{thm:ADT_multicast} Given a linear deterministic single-source $D$-
sink multi-cast wireless network, $\forall \ \epsilon > 0$,  the $\epsilon$-error
capacity $C$ of such a network is given by,
\begin{eqnarray}
\label{eq:ThmLinDetNet} C = \min_{j=1,2,..,D} \ \min_{\omega  \in \ \Omega_j}
\mathrm{rank}(G_{\omega}).
\end{eqnarray}
where $\Omega$ is the set of all cuts between the source and destination $j$. A
strategy utilizing only linear transformations at the relays is sufficient to
achieve this capacity. \ethm

\section{Min-Cut Equals Max Diversity} We begin with a result
applicable to gaussian networks.  \bdefn We define the value
$M_\omega$ of a cut $\omega$ as the number of edges crossing over
from the source side to the sink side across the cut. We refer to
the value of the min-cut as simply the min-cut. \edefn

\bthm \label{thm:mincut} Consider a multi-terminal fading network
with nodes having multiple antennas with each edge having iid
Rayleigh-fading coefficients. The maximum diversity achievable for
any flow is equal to the min-cut between the source and the sink
corresponding to the flow. Each flow can achieve its maximum
diversity simultaneously. \ethm

\bpf We will distinguish between two cases. \vspace{0.05in}

\emph{Case I: Network with single antenna nodes}

\vspace{0.05in}

Choose a source $S_i$ and sink $D_j$. Let $\mathbb{C}_{ij}$ denote the set of all
cuts between $S_i$ and $D_j$.

From the cutset bound \cite{CovTho} on DMT \cite{ZheTse}, \beqan d({0}) & \leq &
\min_{\omega   \in \ \Omega_{ij} } \{d_{\omega}(0) \} =  \min_{\omega   \in \ \Omega_{ij} } \{M_{\omega} \}\\
\Rightarrow d(0)&\leq&M \eeqan

It is now sufficient to prove that diversity order of $M$ is achievable. Let us
first consider the case when there is only one flow.

By the Ford-Fulkerson theorem \cite{FordFulk}, the number of edges
in the min-cut is equal to the maximum number of edge disjoint
paths between source and the destination. Schedule the network in
such a way that each edge in a given edge disjoint path is
activated one by one. Repeat for all the edge disjoint paths. Thus
the same data symbol is transmitted through all the edge disjoint
paths from $S_i$ to $D_j$. Let the number of edges in the $i$-th
edge disjoint path be $n_i$. The $j$th edge in the the $i$th edge
disjoint path is denoted by $e_{ij}$ and the associated fading
coefficient be $h_{ij}$. The activation schedule can be
represented as follows: Activate each of the following edge
individually in successive time instants: $e_{11}, e_{12}, \cdots,
e_{1{n_1}},e_{21}, \cdots, e_{2{n_2}}, \cdots, e_{M1}, e_{M2},
\cdots, e_{M{n_M}}$. Now define $h_i := \prod_{j=1}^{n_i}h_{ij}$
to be the product fading coefficient on the $i$-th path. Let the
total number of time slots required be $N = \Sigma_{i=1}^{M}n_i$.

With this protocol in place, the equivalent channel seen from the
source to the destination has channel matrix

\beqan H & = & \left[
\begin{array}{cccccc}
        h_1         &   0     & \hdots           &0\\
        0           & h_2     &       \hdots       &0\\
        \vdots      &  \vdots    &\ddots & \vdots\\
        0           & 0  & \hdots           & h_M
        \end{array}
        \right]
\eeqan

%Let $u_{ij}$ be defined as: $\rho^{-u_{ij}} := |h_{ij}|^2 $.
%
%If $d_H(r)$ is the outage exponent for this channel,
%
%\beqan \rho^{-d_H(r)} & \doteq &  Pr\{ \Sigma_{i = 1}^{m}
%\log(1+|h_i|^2) \leq rlog\rho\} \\
%& = & Pr\{ \Sigma_{i = 1}^{m} \log(1+\prod_{j=1}^{n_i}|h_{ij}|^2)
%\leq r\log\rho\} \\
%& \doteq & Pr\{ \Sigma_{i = 1}^{m}
%\log(1+\rho^{(1-\Sigma_{j=1}^{n_i}u_{ij})}) \leq r\log\rho\} \\
%& \doteq & Pr\{ \Sigma_{i = 1}^{m}
%\log(1+\rho^{(1-\Sigma_{j=1}^{n_i}u_{ij})}) \leq \log\rho^r\} \\
%& \doteq & Pr\{ \prod_{i = 1}^{m} (\rho^{(1-\Sigma_{j=1}^{n_i}u_{ij})^+}) \leq
%\rho^r\} \eeqan
%
%Following the same lines of arguments as in \cite{ZheTse},
%
%\beq d_H(r) = \inf_{\mathcal{A}} \Sigma_{i=1}^{m}\Sigma_{j=1}^{n_i}u_{ij} \eeq
%
%where
%
%\beq \mathcal{A} = \{ u_{ij}: \Sigma_{i=1}^{m}(1- \Sigma_{j=1}^{n_i}u_{ij})^+ \leq r
%\} \eeq
%
%Let $\Sigma_{j=1}^{n_i}u_{ij} = u_i$. Then,
%
%\beqan
%d(r) & = & \inf_{\mathcal{A}'} \Sigma_{i=1}^{m}u_i \\
%\text{where } \mathcal{A}' & = & \{u_{i}: \Sigma_{i=1}^{m}(1-
%u_{i})^+ \leq r \} \\
%\Rightarrow d_H(r) & = & m-r \eeqan

This matrix is exactly of the structure in Lemma~\ref{lem:Diversity} except that
there are $M$ product Rayleigh coefficients. However, it can be shown that the
diversity remains unchanged. The noise matrix also obeys the conditions of
Lemma~\ref{lem:Diversity}. Thus the maximum diversity of $M$ can be achieved.

When there are multiple flows in the network, we simply schedule
the data of all the flows in a time-division manner. This will
entail further rate loss - however, since we are interested only
in the diversity, we can still achieve each flow's maximum
diversity simultaneously.

\vspace{0.1in}

\emph{Case II: Network with multiples antenna nodes}

\vspace{0.1in}

In the multiple antenna case, we regard any link between a $n_t$
transmit and $n_r$ receive antenna as being composed of $n_tn_r$
links, with one link between each transmit and each receive
antenna. Note that it is possible to selectively activate
precisely one of the $n_tn_r$ Tx-antenna-Rx-antenna pairs by
appropriately transmitting from just one antenna and listening at
just one Rx antenna. The same strategy as in the single antenna
case can then be applied to achieve this diversity in the network.
\epf

\section{Degrees of Freedom of Single Source Single Sink Networks}

In this section, we present a max-flow min-cut type theorem for
evaluating the degrees of freedom in single-source single-sink
networks:

\bthm \label{thm:DOF_ss} Given a single-source single-sink
gaussian wireless network, with independent fading coefficients
having an arbitrary density function,  the DOF of the network is
given by
\begin{eqnarray}
\label{eq:ThmLinDetNet} D = \min_{\omega \ \in \ \Omega}
\mathrm{rank}(H_{\omega}) \text{   with probability one. }
\end{eqnarray}

An amplify-and-forward strategy utilizing only linear transformations at the relays
is sufficient to achieve this DOF. \ethm

\bnote If we assume that the channel coefficients have a time
variation, then we can show that the degrees of freedom is
identically equal to the value shown in the theorem, by coding
across the time variation. \enote

\bpf The proof proceeds as follows: \ben \item First, a converse
for the DOF is provided using simple cutset bounds. \item Then, we
convert the gaussian network into a deterministic network with the
property that the cutset bound on DOF for the gaussian network is
the same as the cutset bound on the capacity of the deterministic
network. \item We then characterize the zero-error-capacity of the
linear deterministic wireless network. \item Finally, we convert
the achievability result for the deterministic network into an
achievability result for gaussian network, which matches the
converse. \een \epf

\subsection{The Converse}
We first provide a simple converse on the degrees of freedom of a single source
single sink network.

\blem \label{lem:ss_converse} Given a single-source single-sink network, the DOF is
upper bounded by the DOF of every cut: \beqan \label{eq:ThmLinDetNet} D & \leq &
\min_{\omega \ \in \ \Omega} \mathrm{rank}(H_{\omega})\eeqan where $H_{\omega}$ is
the matrix corresponding to the cut $\omega$.  \elem

%\bpf The proof is rather straightforward. Consider a cut $\omega \ \in \Omega$
%between the source and destination. The relation between input and output of the cut
%can be written as \beqn Y_{\omega^c} = H_{\omega} X_{\omega} \eeqn. The DOF for this
%channel is given by $r_{\omega} := \text{rank} (H_{\omega})$. Now the DOF of the
%entire network is upper-bounded by the DOF for this particular cut by Lemma~\ref{}.
%This completes the proof. Henceforth, we call the RHS in the converse as $\gamma$,
%i.e., \beqn \gamma := \min_{\omega \ \in \ \Omega} \mathrm{rank}(H_{\omega}) \eeqn
%\epf

\subsection{Conversion to Linear Deterministic Network}
In this subsection, we convert the wireless gaussian network to a equivalent linear deterministic
network.\footnote{It must be noted that the conversion to deterministic network used here is different from that
used in \cite{ADT2} and \cite{CadJafSha}.} We use the term "equivalent" to signify that the DOF of the gaussian
network and the capacity of the deterministic network are the same. In order to get the equivalent deterministic
network, we proceed as follows:

Let the the fading coefficients on the $N$ edges in the gaussian
network be $h_1,h_2,...,h_N$. We first consider a finite field
network with the same graph as the original gaussian network. We
take $q$, the vector length in the deterministic network to be
equal to the maximum number of antennas of any node in the
gaussian network. For nodes with antennas less than $q$, we leave
the remaining nodes un-connected. However, we still need to decide
the finite field size, $p$, and a finite field coefficient on each
edge. Given a finite field size, we need $N$ maps,
$\psi_i,i=1,2,...,N$ that convert the gaussian fading coefficients
into finite field coefficients. Let $\xi_i := \psi_i(h_i)$ denote
this mapping.

In order to obtain these coefficients and the finite field size, we require further
conditions. In particular, we will require the finite field network to have at least
the same capacity as the upper bound on the gaussian network. We recognize the
similarity between the capacity equation in Theorem~\ref{thm:ADT_ss} and DOF terms
in Lemma~\ref{lem:ss_converse} and require that cut by cut, the rank of the transfer
matrix on deterministic network be no less than the rank on the gaussian network.
Before assigning values to $\xi_i$, we will treat them as formal variables.

Consider a cut $\omega$ in the gaussian network. We want the $\text{rank}(G_\omega)
\geq \text{rank}(H_\omega)$ for this cut. To do this, let $r_\omega := \text{rank}
(H_\omega)$ be the DOF of the cut in the gaussian network. Then there exists a
$r_\omega \times r_\omega$ sub-matrix of the $H_\omega$ which has non-zero
determinant. Let us call this matrix as $H^{'}_\omega$. Consider the same cut on the
deterministic network and find the same $r_\omega \times r_\omega$ sub-matrix
$G^{'}_\omega$ corresponding to the transfer matrix $G_\omega$. Now consider the
determinant of the matrix $G^{'}_\omega$. The determinant is a polynomial in several
variables $\xi_i, i=1,2,..,N$ with rational integer coefficients. Let us call this
polynomial as $f_\omega(\xi_1,\xi_2,..,\xi_N)$.

This polynomial is not identically zero as a polynomial over
$\mathbb{Q}$, since in that case even the substitution $\xi_i =
h_i$ will lead to a zero value, making the determinant zero even
for the gaussian case, which is clearly a contradiction. Therefore
we have that $f_\omega$ is a non-zero polynomial. We also have an
observation that the degree of $f_\omega$ in each of the variable
$\xi_i$ is at-most one. We want a field $\mathbb{F}_p$ and an
assignment to $\xi_i$ that makes the $f_\omega$ non-zero over the
chosen field. For any given cut, this can be easily done. However
we want to do this simultaneously for all cuts. To do so, we will
employ the following lemma, proven easily using elementary
algebra:

\blem \label{lem:poly} Given a polynomial $f(\xi_1,\xi_2,...,\xi_N)$ with integer
coefficients, which is not identically zero, there exists a prime field
$\mathbb{F}_p$ with $p$ large enough, such that the polynomial evaluates to a
non-zero value at least for one assignment of field values to the formal variables.
\elem

\vspace*{0.1in}

Now consider the polynomial \beqa f(\xi_1,\xi_2,..,\xi_N) :=
\prod_{\omega  \in \Omega} f_\omega(\xi_1,\xi_2,..,\xi_N) \eeqa
Now, the polynomial $f$ is non-zero since it is a product of
non-zero polynomials $f_\omega$ and the degree of $f$ in any of
the variables is at-most $|\Omega|$. We want a field
$\mathbb{F}_p$ and an assignment for $\xi_i$ from the field such
that $f$ is nonzero. Using Lemma~\ref{lem:poly}, we have that such
an assignment exists. Let us choose that $p$ and the assignment
that makes $f$ non-zero.  Thus we have a deterministic wireless
network whose capacity is guaranteed to be greater than or equal
to $\gamma$ of the converse.

\subsection{Zero Error Capacity of Deterministic Networks}
We establish the zero error capacity of deterministic wireless networks. We have the
following definition

\bdefn \cite{Shannon} The zero error capacity is defined as the supremum of all
achievable rates such that the probability of error is exactly zero. \edefn

\bthm \label{thm:ZEC} The zero error capacity of a single source single sink deterministic wireless
network is equal to \beqan \label{eq:ZEC} C_{ZE} & = &  \min_{\omega \ \in \ \Omega}
\mathrm{rank}(G_{\omega}) \eeqan This capacity can be achieved using a linear code
and linear transformations in all relays. \ethm
\bpf We will prove this theorem using the $\epsilon$ error
capacity result from Theorem~\ref{thm:ADT_ss}.   We will assume
the field $\mathbb{F}$ appearing in the theorem to be the finite
field $\mathbb{F}_p$ of size $p$ where $p$ is the prime previously
identified. From the achievability result in the proof of
Theorem~\ref{thm:ADT_ss}, we have that given any $\epsilon > 0$
and rate $r < C$, there exists a block-length $T$, linear
transformations $A_j, j=1,2,...,M$ of size $qT \times qT$ used by
all relays and a code book $\mathcal{C}$ for the source, such that
the probability of error is lesser than or equal to $\epsilon$.
Each codeword $X_i \in \mathcal{C}$ is a $qT \times 1$ vector that
specifies the entire transmission from the source. Let
$X_1,...,X_{|\mathcal{C}|}$ be the codewords.

Let us assume that the sink listens for a duration $T^{'} \geq T$
in general to account for the presence of paths of unequal lengths
in the network between source and sink, (for large $T$, we would
have $\frac{T^{'}}{T} \rightarrow 1$, so this does not affect rate
calculations). The transfer equation between the source and the
destination vectors are specified by: $ Y = GX$ since all
transformations in the network are indeed linear. Here $G$ is a
$qT^{'} \times qT$ matrix, $X$ is a $T$ length transmitted vector,
and $Y$ is a $T^{'}$ length vector.

Now, given that a vector $X_i$ is transmitted, either the decoder
always makes an error or never makes error because the channel is
a deterministic map $Y_i=GX_i$. Let $P_e^{i}$ be the probability
of error conditioned on transmitting the $i$-th codeword. Then
$P_e^{i} \in \{ 0,1 \} $ and the average codeword error
probability \beqan P_e = \frac{1}{|\mathcal{C}|} \sum_{i=1}^{
|\mathcal{C}|} P_e^{i} \ \leq \  \epsilon & \Rightarrow &
\sum_{i=1}^{ |\mathcal{C}|} P_e^{i} \leq \epsilon |\mathcal{C}|
\eeqan This means that at least $(1-\epsilon)|\mathcal{C}|$
codewords have zero probability of error. Therefore if we choose
only these $(1-\epsilon)|\mathcal{C}|$ codewords as an expurgated
code-book $\mathcal{C}^{'}$, then the code-book has zero
probability of error under the same relay matrices and decoding
rule. The rate of the codebook is however $\bar{r} = r -
\frac{\log (1-\epsilon)}{T}$.  Let $\delta = \frac{\log
(1-\epsilon)}{T}$ be the rate loss and therefore, the expurgated
code-book has negligible rate loss as $T$ becomes large. Now, we
have established a zero error codebook of rate $r - \delta$. By
choosing $r$ arbitrarily close to $C$ and $T$ large, we get that
indeed $C_{ZE} = C$.

However, the code ${\cal C}^{'}$ like the code ${\cal C}$ used in \cite{ADT1}, is a
non-linear code. We obtain a linear code by utilizing the following technique: Since
there is a zero error code for rate $\bar{r}$, it means that the transfer matrix $G$
has rank at least $\bar{r}T$ and therefore that there is a sub-matrix $G^{'}$ of
size $\bar{r}T \times \bar{r}T$, which is full rank. If we communicate only on these
$\bar{r}T$ dimensions we can obtain the transfer matrix $G^{'}$. Thus we get a
linear zero error code of rate $\bar{r}$. \epf \vspace*{-0.05in}
\subsection{Achievable DOF in Gaussian Networks}
In this sub-section, we will lift the zero-error-capacity achievability result from
deterministic networks to determine an achievable DOF for gaussian networks.

In the achievability for capacity of deterministic networks, the
relays performed matrix operations $A_i$ on received vectors for
$T$ time durations.  Since each received vector is of size $q$,
the matrix $A_i$ is of size $qT \times qT$. Now we use the same
strategy for the gaussian network, i.e., all relays use the same
matrices $A_i$ that they used in the deterministic network. This
makes sense, since in a prime finite field $\mathbb{F}_p$, all
field elements are integers modulo $p$. Therefore the matrices
$A_i$ can also be interpreted as matrices over $\mathbb{C}$. This
strategy yields a effective channel matrix $H$, i.e., $Y = HX +
W$.

It is sufficient to prove that $H$ has $\text{rank}(H) \geq \bar{r}T$ since DOF is
equal to $\text{rank}(H)$. To do so, we first establish that there exists an
assignment of $h_i$ such that $\text{rank}(H) \geq \bar{r}T$.

Let us consider the same $\bar{r}T \times \bar{r}T $ sub-matrix
$H^{'}$ by deleting rows and columns in the same way that $G^{'}$
was obtained from $G$. We have that $\det (H^{'})$ is a
multi-variate polynomial in $h_i, i=1,2,...,N$, if we treat $h_i$
as formal variables. Now this polynomial has integer coefficients
and therefore can be treated as a polynomial over any finite
field, in particular over the finite field $\mathbb{F}_p$. Over
$\mathbb{F}_p$, we know that this polynomial is a non-zero
polynomial, since the assignment of $h_i = \xi_i$ gives a non-zero
value. It follows that this polynomial is nonzero, even when
viewed as a polynomial over the integers.  Since $\mathbb{C}$ is
algebraically closed, we have that any non-zero polynomial must
have a assignment of variables in $\mathbb{C}$ that gives non-zero
value to the polynomial. Using this assignment for $h_i$ gives us
that $\det (H^{'}) \neq 0$ and thereby $H$ has $\text{rank}(H)
\geq \bar{r}T$.

We have the following lemma:

\blem Consider a multi-variate polynomial $f$ in several variables $h_i,
i=1,2,..,N$. Let $h_i$ be independent random variables in $\mathbb{C}$ generated
according to any probability density function. If the polynomial has a non-zero
assignment, then the polynomial is non-zero with probability one. \elem

%\bpf The polynomial have non-zero coefficients for at least one of the variables,
%$h_i = h_1$ (say), when treated as a polynomial over the rational function field
%$\mathbb{F}_p(h_2,h_3,...,h_N)$. Then, conditioned on $h_2,..,h_N$ taking specific
%values, the polynomial becomes a univariate polynomial with some degree. However, a
%polynomial has only finitely many zeros and the random variable $h_1$ has a density.
%Therefore, conditioned on $h_2,..,h_N$ taking specific values, we get that the
%probability that the polynomial takes value zero has probability one or that the
%polynomial is non-zero with probability one. On un-conditioning, we get that the
%polynomial is non-zero with probability one. \epf

Now using the Lemma above along with the fact that we have an assignment for $h_i$
such that $\text{rank}(H) \geq \bar{r}T$ with probability one. Therefore, for
channels with frequency (time) selectivity and coding over multiple frequency (time)
slots, we have that the achievable degrees of freedom is equal to $\gamma - \delta$.
Since DOF is defined as the supremum over all achievable DOF values, we have that
DOF $= \gamma$ or \beqan DOF & = & \min_{\{ \omega \ \in \ \Omega\}}
\mathrm{rank}(H_{\omega}) \eeqan

\subsection{Multi-casting over Gaussian Networks}
%In this section, we study the DOF for multi-casting over gaussian networks.
We state the following Theorem without proof:
 \bthm \label{thm:DOF_multicast} Given a single-source
 $D$-sink multi-cast gaussian wireless network,
 with independent fading coefficients having an arbitrary density function, the DOF of the network is given by
\begin{eqnarray}
\label{eq:ThmLinDetNet} D = \min_{ \{j=1,2,...,D \}} \ \min_{\omega \ \in \ \Omega_j}
\mathrm{rank}(H_{\omega}) \text{   with probability one }
\end{eqnarray}
An amplify-and-forward strategy utilizing only linear
transformations at the relays is sufficient to achieve this DOF.
\ethm
%
%\bpf The proof uses Theorem~\ref{thm:ADT_ss} and follows along
%similar lines and is omitted for brevity. \epf
%
\section{Conclusion}
This paper presented two max-flow min-cut type theorems for
computing diversity and DOF of multi-antenna wireless gaussian
networks. In addition, a connection was established between DOF of
gaussian networks and capacity of deterministic networks
\cite{ADT1} for the single-source single-sink and the multi-cast
case.  Along the way, we proved that the zero error capacity of
deterministic networks is the same as the $\epsilon$-error
capacity. While the exact evaluation of capacity for the simplest
relay networks remains open, approximate high SNR
characterizations can be obtained in closed form, even for
arbitrary relay networks using simple amplify-and-forward
protocols.

\end{document}